\PassOptionsToPackage{hyphens}{url}
\RequirePackage{fix-cm}
\documentclass[sn-mathphys,Numbered]{sn-jnl}

\usepackage{mathrsfs}%
\usepackage[title]{appendix}%
\usepackage{textcomp}%
\usepackage{manyfoot}%
\usepackage{algorithmicx}%
\usepackage{algpseudocode}%

\usepackage{framed}
\usepackage{seqsplit}
\usepackage{float}
\usepackage{subfigure}
\usepackage{listings,amsfonts}
\usepackage{amsmath,pifont}
\usepackage{xspace}
\usepackage{bbding}
\usepackage{balance}
\usepackage{hyperref,endnotes}
\usepackage{booktabs}
\usepackage{array}
\usepackage{algorithm}
\usepackage{multirow,makecell}
\usepackage{enumitem}
\usepackage[normalem]{ulem}
\usepackage{graphicx}
\usepackage{amsthm}
\usepackage{multirow}    					
\usepackage{hhline} 
\usepackage{xcolor}
\usepackage{stfloats}
\usepackage{breakurl}
\usepackage[justification=centering]{caption}
\usepackage{soul}
\sethlcolor{white}

\usepackage{amssymb}
\usepackage{pifont}

\newcommand{\tool}{\textsc{WalletRadar}}

\newcommand{\one}{({\em i}\/)\xspace}
\newcommand{\two}{({\em ii}\/)\xspace}
\newcommand{\three}{({\em iii}\/)\xspace}
\newcommand{\four}{({\em iv}\/)\xspace}
\newcommand{\five}{({\em v}\/)\xspace}
\newcommand{\six}{({\em vi}\/)\xspace}

\begin{document}

\title{{\tool}: Towards Automating the Detection of Vulnerabilities in Browser-based Cryptocurrency Wallets} 

\author[1]{\fnm{Pengcheng} \sur{Xia}}\email{xpc357@bupt.edu.cn}
\equalcont{These authors contributed equally to this work.}

\author[1]{\fnm{Yanhui} \sur{Guo}}\email{yhguo@bupt.edu.cn}
\equalcont{These authors contributed equally to this work.}

\author[1]{\fnm{Zhaowen} \sur{Lin}}

\author*[1]{\fnm{Jun} \sur{Wu}}\email{wujun@bupt.edu.cn}

\author[1]{\fnm{Pengbo} \sur{Duan}}

\author[2]{\fnm{Ningyu} \sur{He}}

\author[3]{\fnm{Kailong} \sur{Wang}}

\author[4]{\fnm{Tianming} \sur{Liu}}

\author[5]{\fnm{Yinliang} \sur{Yue}}

\author[6]{\fnm{Guoai} \sur{Xu}}

\author[3]{\fnm{Haoyu} \sur{Wang}}

\affil*[1]{\orgname{Beijing University of Posts and Telecommunications}, \city{Beijing}, \country{China}}

\affil[2]{\orgname{Peking University}, \city{Beijing}, \country{China}}

\affil[3]{\orgname{Huazhong University of Science and Technology}, \city{Wuhan}, \country{China}}

\affil[4]{\orgname{Monash University}, \city{Melbourne}, \country{Australia}}

\affil[5]{\orgname{Zhongguancun Laboratory}, \city{Beijing}, \country{China}}

\affil[6]{\orgname{Harbin Institute of Technology}, \city{Shenzhen}, \country{China}}

\abstract{
Cryptocurrency wallets, acting as fundamental infrastructure to the blockchain ecosystem, have seen significant user growth, particularly among browser-based wallets (i.e., browser extensions). However, this expansion accompanies security challenges, making these wallets prime targets for malicious activities. Despite a substantial user base, there is not only a significant gap in comprehensive security analysis but also a pressing need for specialized tools that can aid developers in reducing vulnerabilities during the development process. 
To fill the void, we present a comprehensive security analysis of browser-based wallets in this paper, along with the development of an automated tool designed for this purpose. We first compile a taxonomy of security vulnerabilities resident in cryptocurrency wallets by harvesting historical security reports. Based on this, we design {\tool}, an automated detection framework that can accurately identify security issues based on static and dynamic analysis. Evaluation of 96 popular browser-based wallets shows {\tool}'s effectiveness, by successfully automating the detection process in 90\% of these wallets with high precision. This evaluation has led to the discovery of 116 security vulnerabilities corresponding to 70 wallets. By the time of this paper, we have received confirmations of 10 vulnerabilities from 8 wallet developers, with over \$2,000 bug bounties. Further, we observed that 12 wallet developers have silently fixed 16 vulnerabilities after our disclosure. {\tool} can effectively automate the identification of security risks in cryptocurrency wallets, thereby enhancing software development quality and safety in the blockchain ecosystem.} 

\keywords{Cryptocurrency, Non-custodial Wallets, Browser Extensions, Automated Security Analysis, Vulnerability Detection, Data Leakage}

\maketitle

\section{Introduction}
\label{sec:introduction}

Cryptocurrencies have captured the attention of a large number of investors in recent years due to their potential economic value. According to the report~\cite{cryptoreport}, the number of cryptocurrency owners has crossed the 500 million milestone by the first half of 2023. Such a substantial user base has made the cryptocurrency market highly active. 
As the number of novice investors swells, there is a growing need for user-friendly software to help them engage with the blockchain, leading to the emergence of \textit{cryptocurrency wallets}. Based on whether the credentials are stored with a centralized third party or in the hands of users, wallets can be classified as custodial and non-custodial, respectively. 
The recent collapse of FTX~\cite{ftxclose}, a provider of custodial wallet services, has highlighted the risks of centralized control, 
leading to a surge in interest in non-custodial wallets, which offer full control and enhanced security of digital assets. 
In particular, browser-based non-custodial cryptocurrency wallets are gaining traction due to their immediate accessibility and straightforward interface. For instance, browser-based wallets like Metamask~\cite{metamask}, Phantom~\cite{phantom}, and Coinbase~\cite{coinbase} have achieved significant success, with over one million downloads. These wallets function as browser extensions and facilitate easy account creation with access to a comprehensive range of blockchain functionalities. 

\textit{Tall trees catch much wind.} Although users can fully control their assets with non-custodial wallets, the responsibility of securely managing credentials also falls on them. This heightened responsibility comes with its own set of challenges, as the security landscape of non-custodial wallets is constantly evolving. In recent years, some incidents have revealed vulnerabilities in even the most reputed non-custodial wallets~\cite{slopehack,checkpoint,checkpoint2,blockworknews,trinityhack}. These incidents have disrupted the notion of non-custodial wallets as fully secure, often resulting in financial losses for users. For example, the Slope Wallet incident led to the leakage of 9,231 wallets' private keys and the loss of about \$4.1 million due to a vulnerability in the wallet's handling of sensitive information~\cite{slopehack}. 
Besides, several malicious software specifically targets browser-based wallets~\cite{attack3,attack2,attack1}. They typically exploit vulnerabilities in wallets to access user credentials, aiming to steal their cryptocurrencies. For instance, by accessing the vulnerable local storage of browser-based wallets, the LummaC2 Stealer was able to steal sensitive information from over 60 wallets on 10 browsers including Chrome and Firefox.

Thus, it is urgent to identify the vulnerabilities of non-custodial wallets and prevent the attacks that exploit them. Indeed, several vulnerability detection tools have been developed by the research community. However, they primarily focus on the security analysis of mobile applications. For instance, Li et al.~\cite{li2020android} explored the attack surface of Android cryptocurrency wallets and found that due to flaws in Android system design and careless development, security issues could expose users' private keys and phrases, risking the financial safety of millions. Uddin et al.~\cite{uddin2021horus} developed a semi-automated framework for assessing the security of Android cryptocurrency wallet apps, revealing critical vulnerabilities in key storage and transaction privacy across numerous applications. 

To the best of our knowledge, the vulnerabilities in browser-based wallets have not been systematically investigated and there is also a lack of automated tools for detecting these vulnerabilities. There are still some questions that the blockchain community is unaware of. 
Firstly, \textit{which are the major vulnerabilities affecting these wallets?} Since blockchain techniques and their corresponding wallets are developing rapidly, the vulnerabilities are also continuously evolving, making it difficult to create a comprehensive and up-to-date list. 
Secondly, \textit{how to reliably and automatically detect these vulnerabilities?} Vulnerabilities common in traditional web applications may manifest differently in browser-based wallets, which could diminish the effectiveness of existing detection methods. Moreover, no automated tools are available to detect newly emerging vulnerabilities dedicated to browser-based wallets. 
Lastly, \textit{to what extent do these vulnerabilities exist?} While there are isolated reports of vulnerabilities, a comprehensive understanding remains unclear regarding the characteristics of these vulnerabilities, the security level of these wallets, and the security awareness among their developers.

\textbf{This work.} We take the first step to characterize and detect vulnerabilities in browser-based wallets. By summarizing security reports of security companies and bulletins of wallet providers, we first create a taxonomy of 6 types of vulnerabilities in browser-based wallets (see \textbf{Section~\ref{sec:taxonomy}}), including traditional web vulnerabilities that also appear in traditional websites but often have a new form of manifestation and new emerging vulnerabilities targeted at cryptocurrency wallets that tend to have more severe security impacts. 
To detect these vulnerabilities, we propose a hybrid approach that combines static and dynamic analysis on browser extensions to accurately and automatically detect vulnerabilities in browser-based wallets (see \textbf{Section~\ref{sec:detection}}). 
The evaluation of 96 popular browser-based wallets shows that our framework can operate on 90\% of the wallets automatically with high accuracy (see \textbf{Section~\ref{sec:vul}}). During this evaluation, 70 wallets (73\% of all wallets) were found to have 116 vulnerabilities. 
These findings underscore a concerning trend: numerous developers of these wallets overlook crucial security mechanisms, such as password policy and credential storage, among others. At last, our impact analysis suggests that these vulnerabilities may influence more than 9.2 million wallet users. However, most wallet developers still have not fixed these issues and do not give enough attention to them. Through our vulnerability disclosure, we assisted 20 different wallets in fixing a total of 26 vulnerabilities. This effort was acknowledged by 8 wallet developers and resulted in \$2,000 in bug bounties. 

In summary, we make the following main research contributions in this paper: 
\begin{itemize}
    \item \textit{We take the first step to create a taxonomy of vulnerabilities in browser-based cryptocurrency wallets.} Through a comprehensive survey of existing security reports and a detailed analysis of popular applications, a taxonomy has been created for 6 types of browser-based wallet vulnerabilities, including traditional ones (i.e., clickjacking, cross-site scripting, defective password policy) and new emerging ones (i.e., demonic vulnerability, redundant storage, defective cryptography). 
    \item \textit{An automated vulnerability detection framework is implemented to identify vulnerabilities in browser-based cryptocurrency wallets.} We developed {\tool}, a framework that combines static and dynamic analysis for accurately identifying vulnerabilities in browser-based wallets. The evaluation shows that the framework can be automated on more than 90\% of the wallets and achieves a high accuracy. 
    \item \textit{We systematically characterize vulnerabilities in \seqsplit{browser-based} cryptocurrency wallets.} This work has revealed that the vulnerabilities are prevalent in \seqsplit{browser-based} wallets and the developers lack attention to them. {\tool} has found that 70 out of 96 tested browser-based wallets are vulnerable. The subsequent impact analysis reveals that more than 9.2 million users face the risk of information leakage and financial loss due to these vulnerabilities. We have received confirmations of 10 vulnerabilities from 8 wallet developers, with over \$2,000 bug bounties. Further, we observed that 12 wallet developers have silently fixed 16 vulnerabilities after our disclosure. 
\end{itemize}

\section{Background} 
\label{sec:background} 

\subsection{Cryptocurrency Wallets} 

Cryptocurrencies, originally developed as a component of blockchain's reward mechanism, play a critical role within the blockchain system. The first cryptocurrency Bitcoin~\cite{bitcoinhome} was released in 2009, and to date, there are over 23 thousand cryptocurrencies worldwide~\cite{coinmarketcap}. Following the surge in cryptocurrency popularity in 2017~\cite{bitcoinhype}, individuals have flooded into cryptocurrency markets to acquire or trade cryptocurrencies. For most people without technical experience, using a cryptocurrency wallet is essential to perform transactions on a blockchain platform.  
This wallet, which can be either a software or hardware tool, facilitates the storing and trading of cryptocurrencies by interacting with blockchain ledgers. 
However, unlike conventional wallets that physically store fiat currency, cryptocurrency wallets do not directly store digital assets. 
Since cryptocurrencies inherently exist as transaction data within the blockchain ledgers, the wallet validates the user's cryptocurrency holdings by retrieving the user's transaction information corresponding to their unique addresses (i.e., blockchain accounts). 

According to how keys are stored, cryptocurrency wallets can be generally classified into two categories: \one~\textit{custodial wallets}, which depend on a centralized third party for key storage, and \two~\textit{non-custodial wallets}, which store keys locally. Non-custodial wallets have been gaining popularity due to their enhanced personal security and direct ownership of assets. They also align with the decentralization principle that is at the heart of the cryptocurrency paradigm. 

Non-custodial wallets include browser-based wallets, desktop wallets, and mobile wallets. 
Existing research~\cite{sai2019privacy,hu2021security} has revealed a range of security issues associated with mobile wallets and their operating environments. Browser-based wallets, despite their considerable user base owing to their online accessibility, lack systematic security analysis. 
One such browser-based wallet, MetaMask, has an impressive user base exceeding 10 million. 
The wallet application can be downloaded from the browser extension store and operated locally, with the user data also stored in the local browser, providing users with full control. 
However, browser-based non-custodial wallets including MetaMask are not exempt from security threats, and vulnerability-related incidents are frequently reported~\cite{slopehack,checkpoint2,blockworknews}. 
Therefore, our research primarily focuses on such wallets, aiming to unearth the distinct security issues prevalent in this class of wallets, create a taxonomy derived from these vulnerabilities, and develop methodologies for their identification.

\subsection{General Workflow of a Browser-based Wallet}

\begin{figure}[htbp]
\centering
\includegraphics[width=0.7\linewidth]{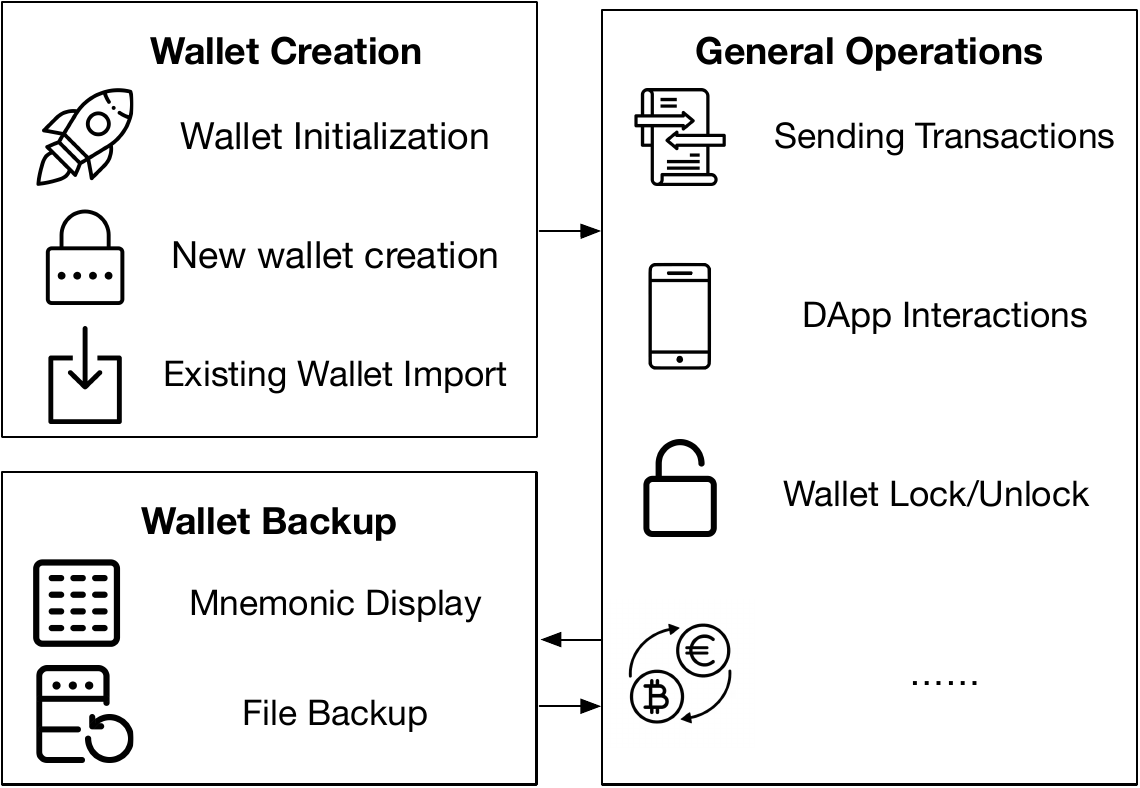} 
\caption{The general workflow of the browser-based cryptocurrency wallets.} 
\label{fig:lifecycle} 
\end{figure}

As shown in Figure~\ref{fig:lifecycle}, the main function of a browser-based wallet usually includes wallet creation, wallet backup, and other general wallet operations. 

\one \textbf{Wallet Creation}: After the wallet's initial launch and traversing the start page, users can opt either to create a brand-new wallet or import an existing one. 
When creating a new wallet, users need to set a password, which is used to unlock the wallet. Then the wallet will generate a pair of keys, i.e., a private key and a public key, and further output the wallet address based on the public key. In addition, the wallet will automatically generate a string of words for the user (called ``mnemonics'') to recover the wallet.
As for integrating an existing wallet, users should provide the original private key or mnemonics and set the wallet password to complete the wallet import. 

\two \textbf{Wallet Backup}: In the wallet backup process, users are firstly required to provide a password to pass the identity authentication. Following this, the wallet will display the mnemonic or private key in plaintext for users to copy and back up elsewhere. Additionally, a few wallets offer methods to back up these sensitive data to files. In such cases, the wallet might require users to provide another temporary password to encrypt the backup file, thereby ensuring its security. 

\three \textbf{General operations}: Wallets may be involved in other general operations, like transaction initiation, and decentralized applications (DApps) interactions. In these operations, wallets serve as a medium to help users communicate with the blockchain in terms of transaction activities.
For security reasons, when it comes to sensitive operations, e.g., wallet backup, the wallet mandates a password-based user authentication, i.e., users are required to give the password to unlock and perform these sensitive operations. 

\section{Vulnerability Taxonomy} 

\subsection{Generating the Taxonomy} 
\label{sec:taxonomy}

To understand and identify vulnerabilities that are inherent to browser-based cryptocurrency wallets, we first propose a taxonomy. 
Specifically, to conduct a comprehensive analysis, we focus on behaviors in each stage of the lifecycle of wallets, as shown in Figure~\ref{fig:lifecycle}. 
In other words, we concentrate on sensitive operations at wallet creation, backup, and unlock. Note that, since the security of blockchain interactions, like transaction processing, primarily lies on the blockchain side, this work does not consider the threats of these activities. 

We aim to divide the vulnerabilities into the following two aspects: 
\one \textit{Traditional web vulnerabilities}. Given that web extensions fundamentally operate on web pages, gathering traditional web vulnerabilities and considering their applicability is required. This encompasses a range of vulnerabilities like injection, XSS, access control flaws, and clickjacking. 
\two \textit{New emerging cryptocurrency wallet vulnerabilities}. Even though the scope of the first type is extensive, vulnerabilities that may have a limited impact on traditional web applications could pose a more serious threat in browser-based wallets. Therefore, inspired by existing literature on mobile wallets~\cite{he2020security,li2020android,uddin2021horus}, this category consists of some specific vulnerabilities like sensitive data management, flaws in the storage process, and improper use of cryptographic methods, which could cause sensitive data leakage or even financial loss in browser-based cryptocurrency wallets. 

\hl{Thus, in addition to sources directly related to browser-wallet security, we also considered literature on general web security to ensure our taxonomy's robustness. This approach allowed us to include insights from both blockchain-specific vulnerabilities and those common in broader web applications, providing a comprehensive view of the security landscape for browser-based wallets. In specific,} we compile a list of the security vulnerabilities in browser-based wallets from related research~\cite{hu2021security,sai2019privacy}, best-industrial practice guidelines~\cite{appsealing,algorand,OWASPtop} and security reports~\cite{metamasksec,ledger,certik}. \hl{We systematically reviewed each source to gather information on vulnerabilities specific to browser-based wallets. This involved assessing the characteristics and impacts of each vulnerability to decide its relevance to browser-based wallets. For accuracy and to ensure no detail was overlooked, two authors independently checked the classifications. The vulnerabilities are finally divided into two categories we mentioned above, as shown in} Table~\ref{tab:wallet vulnerabilities}.

\begin{table}[h]
\caption{The vulnerabilities of browser-based cryptocurrency wallets.} 
\centering
\label{tab:wallet vulnerabilities} 
\begin{tabular}{@{}ll@{}}
\toprule
Category                  & Description                                \\ \midrule
Clickjacking              & Overlaying phishing pages to trick users.  \\
XSS                       & Injecting harmful scripts into pages.      \\
Defective Password Policy & Permitting weak, easily cracked passwords. \\
Redundant Storage         & Unnecessary storage of sensitive data.     \\
Demonic Vulnerability     & Insecure caching of sensitive keys.        \\
Defective Cryptography    & Use of weak cryptographic methods.         \\ \bottomrule
\end{tabular}
\end{table}

\subsection{Traditional Web Vulnerabilities}

\subsubsection{Clickjacking} 
\label{sec:clickjacking} 
Clickjacking is a vulnerability that employs visual deception~\cite{calzavara2020tale, huang2012clickjacking, wu2016analysis}. In the context of cryptocurrency wallets, attackers overlay a transparent wallet homepage, extracted from the target browser-based wallet, on a phishing webpage carefully crafted by them. When careless users interact with the deceptive page, their wallets will be manipulated. This can lead to unauthorized fund transfers or sensitive data leakage. 
In contrast to the traditional clickjacking defenses, which incorporate headers like Content Security Policy (CSP) and X-Frame-Options to restrict browsers from rendering embedded pages in responses, browser extensions rely on their configuration files to control access from external websites. Thus, if the wallet's main HTML page is listed under the ``web\uline{ }accessible\uline{ }resources" configuration in the ``manifest.json'' file, external pages can access the main page, potentially introducing clickjacking vulnerability.

\subsubsection{Cross-Site Scripting (XSS)} 
\label{sec:xssintro}
XSS attacks exploit web functions that render dynamic content~\cite{gupta2017cross,kumar2022comprehensive,kumar2022enhanced}. Attackers usually insert malicious JavaScript code into HTML pages to manipulate the rendered content. When it comes to browser-based cryptocurrency wallets, there are two notable impacts, i.e., \textit{automatic page manipulation} and \textit{unauthorized access to local storage} (such as localStorage and indexedDB). While the former can lead to unauthorized fund transfer, the latter could result in the leakage of encrypted data. 

Web extensions usually provide users with dynamic notifications, automatic redirection, and other features through the HTML Document Object Model (DOM). After the HTML DOM is loaded, cryptocurrency wallets often alter the DOM to specific page contents. If these alterations involve sensitive functions, they may result in DOM-based XSS vulnerabilities. 

\subsubsection{Defective Password Policy} 
Requiring users to set a complex password is crucial. In browser-based cryptocurrency wallets, all the information in users' wallets is encrypted and stored locally. If a weak password is adopted, attackers may attempt brute force attacks, causing users to lose control of their wallets. A CheckPoint report~\cite{checkpoint} points out that when attackers obtain locally stored cryptographic wallet information through malicious means, they can brute-forcedly try 95 passwords per second on a 4-core Intel Core i7 CPU, which is sufficient to exploit a weak 6-digit password.

\subsection{New Emerging Cryptocurrency Wallet Vulnerabilities} 

\subsubsection{Redundant Storage} 
\label{sec:redundant} 
Redundant storage vulnerabilities arise when wallets store sensitive information or related intermediary processing results in local storage. This can significantly lower the barrier for attackers to access sensitive information. 
It usually occurs in the locking or unlocking stage of cryptocurrency wallets. If wallets store intermediate decryption results in the browser, attackers who can exploit this data (e.g., through an XSS attack) can reverse-engineer and reproduce the decryption sequence. 
To fully understand this vulnerability, consider a typical wallet unlocking process as an example. 

\one \textbf{Data Retrieval}. The wallet receives the password entered by the user and retrieves the locally stored encrypted wallet data. 

\two \textbf{Key Generation}. Using passwords directly for wallet authentication can make the system vulnerable to brute-force and rainbow table attacks, as it heavily relies on password strength. A common practice for authentication is to create a decryption key from the password using methods with hash iterations (called Password-Based Key Derivation Function, PBKDF), which adds an extra layer of complexity and security~\cite{turan2010recommendation,visconti2019examining}. To generate a decryption key, the wallet either uses the user's raw input for hashing iterations or initially hashes the user's input, then employs the obtained password hash to further hashing iterations (with another hashing method). 

\three \textbf{Data Decryption}. With the generated key, the wallet tries to decrypt the encrypted data. If it succeeds and unveils the plain wallet data, the user is navigated to other pages for the following operations. 

During our observations, we found that some wallets would carelessly store sensitive data, such as the hash calculated from the password (step {\em ii}), in local storage even after unlocking. 
Note that the implementation of the unlocking process is transparent, which can be easily obtained by auditing the front-end JavaScript files.
Thus, after obtaining this hash, they just need to reproduce the iterative hash process to generate the decryption key (step {\em ii}) and then use the key to unlock the wallet (step {\em iii}). The decrypted wallet data usually contains mnemonics or private keys, which gives attackers full control of the wallet.

\subsubsection{Demonic Vulnerability} 
Demonic vulnerability is another storage-related vulnerability. Unlike the redundant storage vulnerability, which results from the inherent wallet design, the demonic vulnerability is rooted in browser caching mechanisms, leading to the unintentional local caching of sensitive data. 
According to the report~\cite{metamask_demonic}, the early versions of Metamask held mnemonics in the HTML ``textarea'' tag when importing wallets. As an inherent browser mechanism, browsers are designed to cache textual data from active tabs to preserve the current state of the page, allowing for faster access and retrieval later. Thus, such sensitive information would be saved to the local disk due to the caching mechanism. Given the importance of mnemonics in the aspect of cryptocurrency wallets, this is a significant security concern. 
Numerous wallets in the market employ similar implementations to display sensitive data. As these implementations can be found in various functions, including wallet imports, wallet creation, mnemonic display, and private key display, the potential impact of this kind of vulnerability is extensive.

\subsubsection{Defective Cryptography} 
\label{sec:defective_crypto} 

Cryptographic algorithms, central to the functionality of browser-based wallets, are pivotal for sensitive operations like wallet creation and identity authentication, as highlighted in \S\ref{sec:taxonomy}. This reliance on cryptography underlines the importance of their proper implementation in securing wallet data, which is the reason we put it in the ``new emerging cryptocurrency wallet'' category. In creating our taxonomy, we noticed a knowledge gap among some wallet developers regarding the optimal adoption of these algorithms, leading to significant cryptography-related issues. 

One common issue is the insufficient iterations of the PBKDF2 algorithm (mentioned in \S\ref{sec:redundant}). We observed wallets employing as few as 100 or $5,000$ iterations, falling short of the recommended $10,000$ and optimal $310,000$ rounds. This inadequate iteration count makes wallets more vulnerable to brute-force attacks, a risk that escalates with weak passwords. Additionally, the choice of encryption patterns is critical. The use of AES-CBC mode, for example, poses risks due to its lack of integrity checks. A more secure alternative like AES-GCM, which offers both confidentiality and integrity, would be preferable.

\definecolor{lightgray}{rgb}{.9,.9,.9}
\definecolor{darkgray}{rgb}{.4,.4,.4}
\definecolor{purple}{rgb}{0.65, 0.12, 0.82}

\lstdefinelanguage{JavaScript}{
  keywords={typeof, new, true, false, catch, function, return, null, catch, switch, var, if, in, while, do, else, case, break},
  keywordstyle=\color{blue}\bfseries,
  ndkeywords={class, export, boolean, throw, implements, import, this},
  ndkeywordstyle=\color{darkgray}\bfseries,
  identifierstyle=\color{black},
  sensitive=false,
  comment=[l]{//},
  morecomment=[s]{/*}{*/},
  commentstyle=\color{red}\ttfamily,
  stringstyle=\color{purple}\ttfamily,
  morestring=[b]',
  morestring=[b]"
}

\lstdefinestyle{styleHTML}{
    language=HTML,
	extendedchars=true,
	basicstyle=\footnotesize\ttfamily,
    commentstyle=\color{magenta},
    keywordstyle=\color{blue},
	showstringspaces=false,
	showspaces=false,
	numbers=left,
	numberstyle=\footnotesize,
	numbersep=5pt,
	tabsize=2,
	breaklines=true,
	showtabs=false,
	captionpos=b,
	xleftmargin=0.5cm,
}

\lstdefinestyle{styleJS}{
    language=JavaScript,
	extendedchars=true,
	basicstyle=\footnotesize\ttfamily,
	showstringspaces=false,
	showspaces=false,
	numbers=left,
	numberstyle=\footnotesize,
	numbersep=5pt,
	tabsize=2,
	breaklines=true,
	showtabs=false,
	captionpos=b,
	xleftmargin=0.5cm,
}

\section{The Design of {\tool}} 
\label{sec:detection}

In this section, we present {\tool}, an automated vulnerability detection framework specifically designed for identifying vulnerabilities in browser-based wallets. 
According to our taxonomy, we aim to detect all six kinds of vulnerabilities. 

We seek to design a hybrid approach that combines static and dynamic analysis. 
On the one hand, clickjacking, XSS vulnerabilities, and defective cryptography in browser-based wallets can be easily identified through an efficient static analysis.
For example, identifying defective cryptography can be done via filtering the signatures of adopted cryptography algorithms. Due to the simplicity of identifying these vulnerabilities, the inherent false positive issue can be reduced as much as possible. 
In contrast, static analysis is not suitable for detecting the other three vulnerabilities. Since requirements and implementations for password policies vary across wallets, conducting dynamic testing on password input is more appropriate. Also, demonic vulnerabilities and redundant storage vulnerabilities require monitoring the dynamically changing webpage information and data in local storage. Therefore, we propose another set of dynamic analysis methods to identify these vulnerabilities. 

\begin{figure}[t]
\centering
\includegraphics[width=0.7\linewidth]{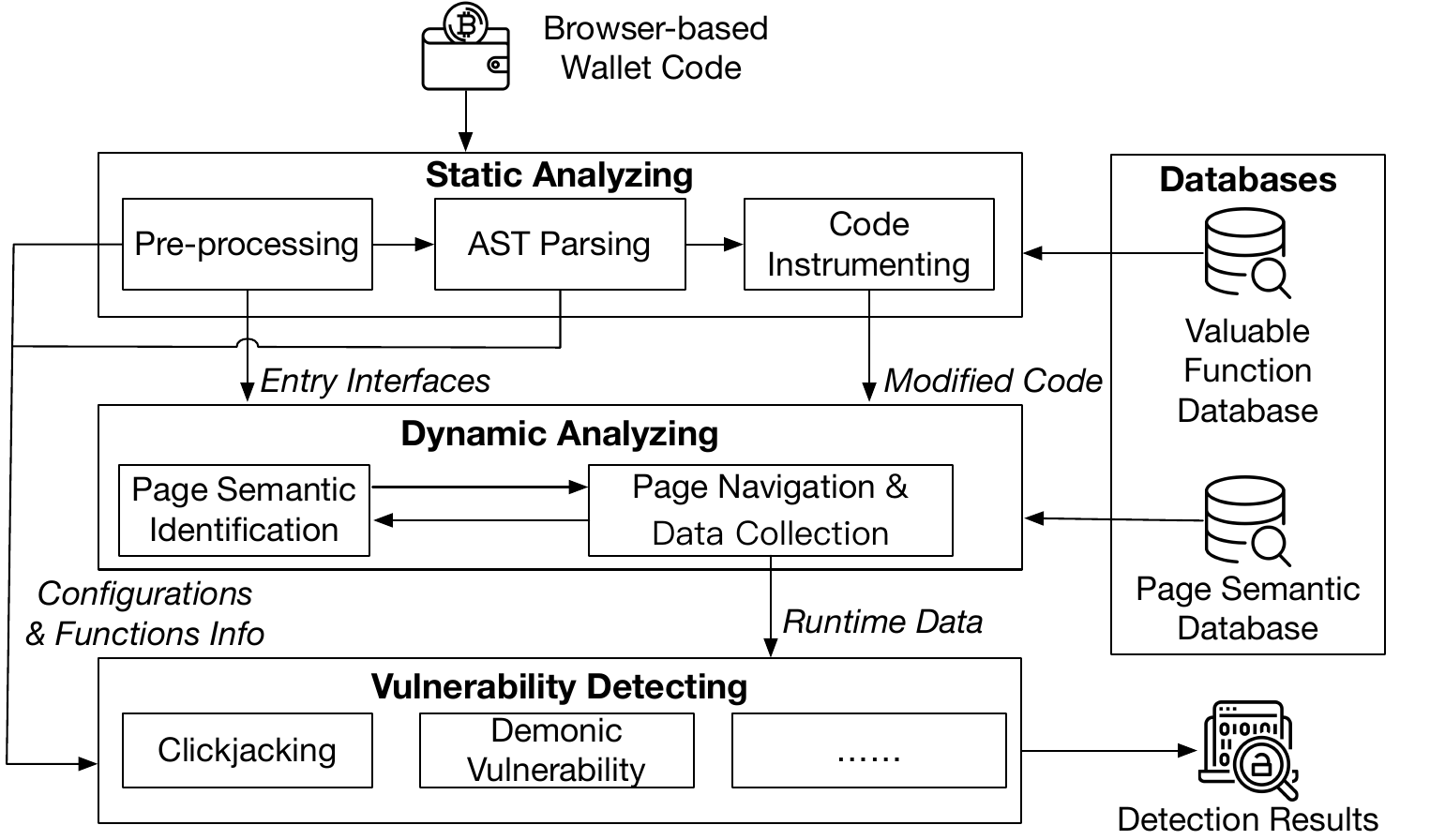} 
\caption{The detection workflow for \tool{}.} 
\label{fig:detection overview} 
\end{figure}

\subsection{Approach Overview} 
\label{sec:approach:overview} 
Figure~\ref{fig:detection overview} provides a general overview of {\tool}. The framework includes three core phases: \one static analyzing, \two dynamic analyzing, and \three vulnerability detecting. 

Specifically, {\tool} takes the code base of a to-be-verified wallet as input.
In the static analyzing phase, {\tool} firstly performs necessary pre-processings. It not only deobfuscates and reformats the given wallet, but also tries to extract entry interfaces of it. Based on the beautified and readable source code, {\tool} parses the corresponding Abstract Syntax Trees (ASTs). With the help of the \textit{valuable function database}, {\tool} can filter the ASTs of suspicious functions out. Meanwhile, {\tool} collects static features, like the locations and some hard-coded parameters of target functions. 
To obtain the runtime data, {\tool} has to conduct the necessary instrumentation before each function invocation. 
In the dynamic analyzing phase, based on the entry interfaces and the instrumented source code, {\tool} dynamically deploys the wallet in our local testing environment. According to the \textit{page semantic database}, {\tool} can simulate the user interactions with pages, and collects the runtime data (like local storage). 
Finally, in the vulnerability detecting phase, according to the data collected from the above two phases, {\tool} can efficiently and effectively identify all six types of vulnerabilities.

\subsection{Static Analyzing} 
\label{sec:approach:static} 
The static analyzing phase is responsible for collecting static features and conducting instrumentation for the following dynamic analyzing phase. 
It consists of three stages, i.e., \textit{pre-processing}, \textit{AST parsing}, and \textit{code instrumenting}. 

\subsubsection{Pre-processing}
\label{sec:approach:static:preprocessing}
As we stated in \S\ref{sec:approach:overview}, pre-processing is mainly responsible for two things: \textit{beautifying}, and \textit{static features extracting}.
Thus, {\tool} firstly tries to deobfuscate and reformat the source code by js-beauty~\cite{jsbeauty}, a well-known and widely adopted formatting tool.
Then, to extract features, {\tool} parses the configuration files (e.g., ``manifest.json'') of the wallets.
To be specific, we focus on the fields related to entry interfaces, like the ``background'' field, which defines scripts or pages that run persistently in the background; ``action'' defines the pages displayed when users open the wallets; ``content\_scripts'' defines scripts that are injected and executed on web pages supported by the wallets; and ``web\_accessible\_resources'' defines resources within the wallet that can be accessed by other web pages. The configuration items will be directly sent to the vulnerability detecting phase for further analysis. 
The extracted possible entry interfaces will be sent to the dynamic analyzing phase. 

\subsubsection{AST Parsing} 
\label{sec:approach:static:ast} 

Based on the implementation of the wallet, AST parsing will filter suspicious functions out and parse them into AST format. 
Specifically, {\tool} invokes Espri\-ma~\cite{esprima}, a widely-used JavaScript parser, to obtain the AST for all functions. Then, according to the \textit{valuable function database}, which is collected from related research~\cite{hu2021security,sai2019privacy} and reports~\cite{certik,metamask}, {\tool} can identify potentially vulnerable APIs as well as their corresponding AST through simple but effective regex matching. \hl{Two types of APIs in the valuable function database are concerned, i.e.,} \textit{cryptographic algorithm functions}, and \textit{DOM manipulating functions.} \hl{The cryptographic algorithm functions deal with sensitive data handling, like generating decryption keys through key derivation functions or decrypting wallet data (e.g., ``AES.decrypt'' and ``crypto.subtle.deriveKey''). On the other hand, DOM manipulating functions are typically used for providing user notifications or facilitating automatic webpage redirections (e.g., ``document.write'' and ``window.location.replace'').}

\hl{To identify corresponding AST of these APIs, for cryptographic algorithm functions, we need to find the accurate locations of them and extract hard-coded parameters in them.} Thus, an extended forward search is initiated to start from the outermost function of the current file, targeting the matched function, and to record the search path encountered during this process. \hl{If certain cryptographic algorithm parameters are hard-coded in the code, they are recorded during this step. This step is essential as it can be challenging to obtain these parameters through dynamic code instrumentation, which is primarily intended for capturing dynamically passed variables within functions. This approach ensures that no critical information of related functions is overlooked, thereby enhancing the accuracy of our analysis.} 
As for DOM manipulating functions, to check whether their data sources can be modified, {\tool} first needs to find their data source functions. Thus, a taint backtracking is performed from the matched function to the data source function on the AST of the source code. Similarly, {\tool} will record the search paths for subsequent vulnerability detection.

\subsubsection{Code Instrumenting}
\label{sec:approach:static:instrumenting}

Code instrumentation is for better collecting runtime data in the following dynamic analyzing phase. 
Therefore, as we mentioned in Section~\ref{sec:approach:static:ast}, two types of functions are kept: \textit{cryptographic algorithm related} and \textit{DOM manipulating related}. 
Since we are concerned with parameters that might be dynamically passed into cryptographic functions, such as keys and ciphertext, and the data flow of DOM functions has already been obtained through AST parsing, we only conduct code instrumentation on cryptographic algorithm related functions. 
Specifically, based on preliminary experimental results, we find that cryptographic algorithm related functions are usually enveloped by another layer of functions, {\tool} instrument the location of this enveloping function for further parameter collection. 
Take a decrypting function under the cryptographic algorithm related category as an example. As we can see from Figure~\ref{fig:code example1}, at Line 13, it invokes the actual AES decrypting function with parameters \texttt{c} and \texttt{b}, where \texttt{b} is passed through argument directly and \texttt{c} is derived from another argument \texttt{a}. When identifying the function (i.e., \texttt{unlock}) that wraps the target function, {\tool} performs a local data flow analysis to locate its parameters (i.e., \texttt{a}) related to the decryption function, establishing such a data dependency relation. 
Then, we instrument a \texttt{collect}, our self-defined function, before the assignment to \texttt{c} at Line 7. 
Consequently, through the instrumented function, we can obtain the values that are passed to the actual decryption function. 

\begin{figure}[htbp]
\centering
\begin{lstlisting}[style=styleJS]
function UnlockExample(x, y, z) {
    function process(temp) {
        ...
    }
    function unlock(a, b) {
      // The code snippet being injected to collect the parameters
      //collect();
      var c = process(a);

      function unlocklog(d) {
        console.log(d);
      }
      const decrypted = CryptoJS.AES.decrypt(c, b); 
    }      
}
\end{lstlisting}
\caption{An example of a decryption function.}
\label{fig:code example1}
\end{figure} 

\subsection{Dynamic Analyzing} 
\label{sec:dynamic} 

This phase is intended to capture critical runtime data from the wallets, like local storage data, function innovation details, and dynamic HTML content, which is impossible to collect by the static analysis. All collected data will be used for the final vulnerability detection.

\subsubsection{Page Semantics Identification} 
\label{subsec:semantics} 

To conduct an effective runtime information collection, {\tool} must first identify the functionalities of the current page, i.e., the semantics of the page. 
As the concrete operating processes of different wallets vary, it is non-trivial to implement a set of general page semantics identification methods. 
Thus, against sensitive operations, like password setting and mnemonic import, we preliminarily conduct a manual investigation to find common pages that these browser-based wallets may share. Then, we build a semantics database based on the page patterns, like specific text keywords and HTML elements. 
Taking advantage of this semantics database, {\tool} is able to efficiently identify the functionality of the current page during the dynamic analysis, and perform the corresponding following operations. 
We then detail the \textit{manual investigation on key pages} and \textit{semantics database build} in the following. 

\begin{table*}[h]

\centering
\caption{The 13 key pages and their functionalities.}
\label{tab:keypage}
\resizebox{0.95\textwidth}{!}{
\begin{tabular}{ccl}
\toprule
Stage                                                                              & Page                                                             & \multicolumn{1}{c}{Functionalities/Page Content} \\ \midrule
\multirow{3}{*}{\begin{tabular}[c]{@{}c@{}}Wallet \\ Initiation\end{tabular}}      & Start Page                                                       & Entry point with welcome information.             \\ 
                                                                                   & Data Collection Reminder                                         & User agreement and privacy policy.                \\ 
                                                                                   & Wallet Creation Preparations                                     & Select wallet creation method.                    \\ \midrule
\multirow{3}{*}{\begin{tabular}[c]{@{}c@{}}New Wallet \\ Creation\end{tabular}}    & Password Setting                                                 & Set wallet access password.                       \\  
                                                                                   & Security Reminder                                                & Displays security best practices.                 \\ 
                                                                                   & Mnemonic Display                                                 & Shows backup mnemonic phrase.                     \\ \midrule
\multirow{2}{*}{\begin{tabular}[c]{@{}c@{}}Existing Wallet \\ Import\end{tabular}} & Import Method Selection                                          & Choose wallet import method.                      \\
                                                                                   & Mnemonic Import                                                  & Wallet recovery via mnemonic.                     \\ \midrule
\multirow{2}{*}{\begin{tabular}[c]{@{}c@{}}General \\ Operations\end{tabular}}     & Home Page                                                        & Main interface with functionalities.              \\ 
                                                                                   & Wallet Unlock                                                    & Unlocking the wallet interface.                   \\ \midrule
\multirow{3}{*}{\begin{tabular}[c]{@{}c@{}}Wallet \\ Backup\end{tabular}}          & Wallet Setting                                                   & Wallet configuration settings.                    \\ 
                                                                                   & \begin{tabular}[c]{@{}c@{}}Password \\ Verification\end{tabular} & Verify password for backup.                       \\ 
                                                                                   & Wallet Backup                                                    & Interface for wallet backup.                      \\ \bottomrule
\end{tabular}
}
\end{table*}

\begin{figure}[htbp] 
\centering
\includegraphics[width=0.8\linewidth]{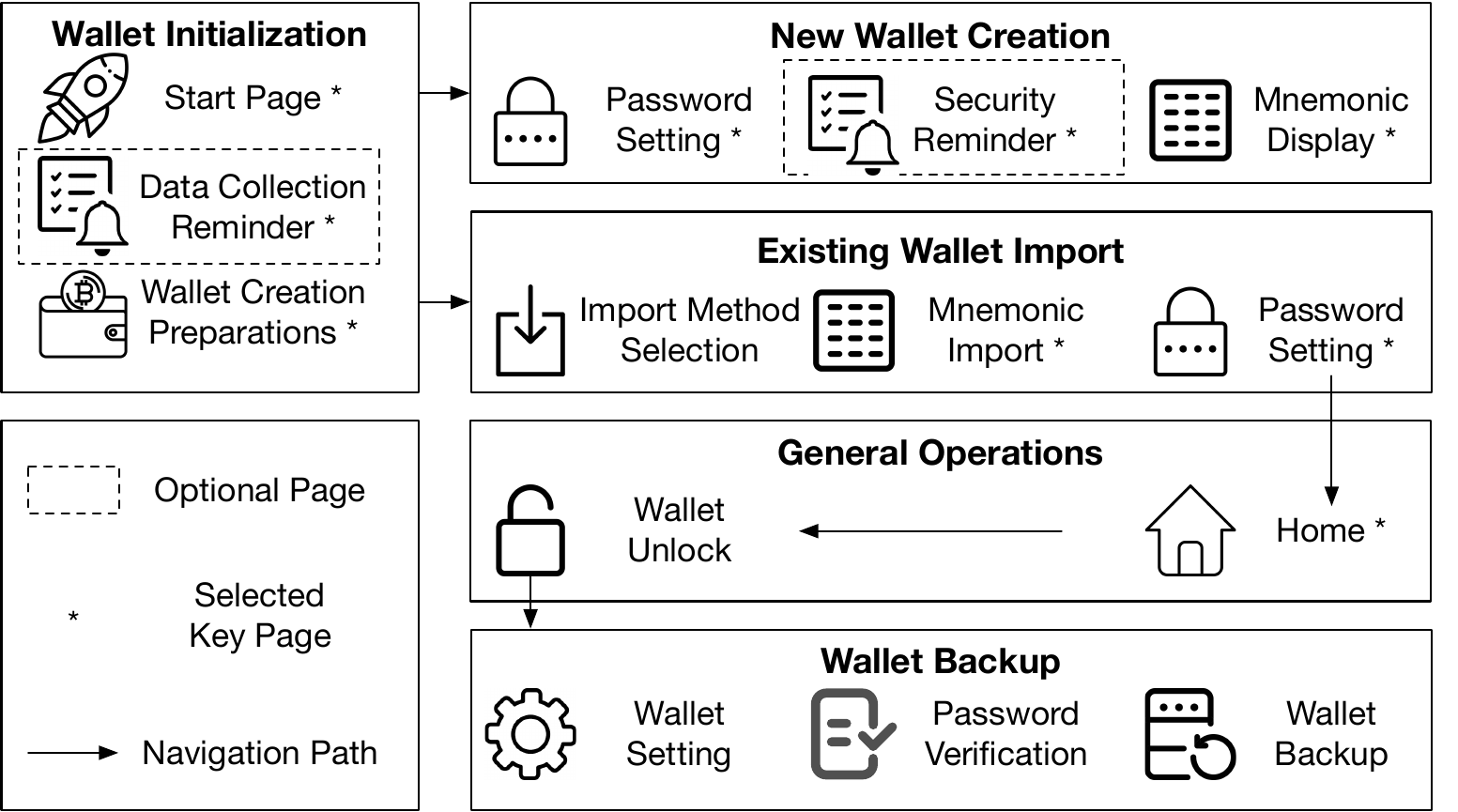} 
\caption{The 13 key pages we select and the navigation path during dynamic analyzing. Note that, the password setting page appears twice.} 
\label{fig:wallet activity} 
\end{figure} 

\textbf{Manual Investigation on Key Pages.} 
The manual investigation process involves examining each wallet's user interface, noting down the standard and sensitive features such as wallet creation, wallet import, password setting, and wallet backup. By cataloging these pages, we were able to define a set of generic templates that represent the core functional pages across different wallets. 
After the analysis, 13 key pages remain that can cover all sensitive operations in a browser-based wallet, as shown in Figure~\ref{fig:wallet activity}. 
These thirteen pages, following the wallet's usage flow, can be divided into five stages: wallet initialization, new wallet creation, existing wallet import, general operations, and wallet backup. The functionalities of these pages are shown in Table~\ref{tab:keypage}. 
As we can see, except for two optional reminder pages, most pages adhere to basic sequential order. In particular, under clearly defined stage sequences, some pages consistently appear after specific ones. 
For instance, when a user clicks the ``Lock" button on the home page, the wallet will navigate to the unlock page. 
Hence, predictable pages are excluded, and the others are selected as pages requiring the construction of semantic features. Note that, during analysis, we find many wallets merge mnemonic import and password setting features on a single page (referred to as ``wallet setup page'') when importing mnemonics, and we add this page to our semantic library. At last, 9 key pages, marked with a ``*'' symbol in Figure~\ref{fig:wallet activity}, have been selected for constructing their semantics.

\textbf{Semantics Database Build.} 
To enable efficient semantics identification on pages, we build a semantics database. 
Specifically, each row of the database is organized as a key and a series of features, where the key is the corresponding functionality, and the features are composed of several metrics of the functionality, including specific text keywords and HTML elements. 
For example, mnemonic import is one of the major functionalities of the wallet setup page. Thus, we first build the keyword features of the functionality. The Term Frequency-Inverse Document Frequency (TF-IDF) method, a statistical measure used in text mining and information retrieval~\cite{qaiser2018text}, is adopted. This method helps to identify how important a word is to a document in a collection or corpus. Applying TF-IDF to 20 popular wallets, we calculate the frequency of words appearing during the mnemonic import. Subsequently, based on our understanding of sensitive operations, we frame the primary keyword semantics of the functionality in an ``Action + Object'' format. This involves selecting specific verbs that capture the essence of the functionality's activities (e.g., ``import'' and ``input''), paired with nouns that represent the subjects of these actions (e.g., ``recovery phrase'' and ``mnemonics''). Under this guideline, the keyword semantics for each functionality comprise a few groups of words or phrases. 
Besides, we also consider interactive HTML elements within the functionalities, such as an input box or 12-24 consecutive input boxes for mnemonic phrase entry during the mnemonic import process. 

Table~\ref{tab:page semantics} displays the keyword semantics we established for the wallet setup page, encompassing five unique keyword groups: two related to the wallet import and three associated with the creation of a new password. A functionality is identified with a specific semantic if it matches at least one keyword or phrase in each group of the functionality and contains the requisite HTML elements. Furthermore, identifying the semantics of a page implies recognizing the semantics of all its functionalities. Following the rules, a page with keywords like ``import'', ``mnemonic'', ``password'', ``enter'', ``repeat'' and corresponding input fields for mnemonics and passwords is identified as a wallet setup page. 

\begin{table}[h] 
\centering 
\small 
\caption{The keyword semantics for the wallet setup page.} 
\label{tab:page semantics} 
\begin{tabular}{@{}cl@{}} 
\toprule 
Group No. & \multicolumn{1}{c}{Semantic Words or Phrases} \\ \midrule
1         & import, input, provide, ...                   \\
2         & recovery phrase, mnemonics, seed phrase, ...   \\
3         & password, credential, PIN, ...                \\
4         & create, enter, type in, ...                   \\
5         & repeat, confirm, verify, ...                  \\ \bottomrule
\end{tabular} %
\end{table}

\subsubsection{Page Navigation \& Runtime Data Collection} 

With the help of the semantics database, {\tool} needs to perform appropriate actions according to the current page's functions and elements in order to traverse the wallet's main features while concurrently collecting runtime data. 
Besides, as we mentioned in \S\ref{subsec:semantics}, pages often display in a certain order. Determining navigation paths for these pages can enhance the efficiency of the dynamic analysis. 
While navigating through the pages, the functions instrumented in the static analyzing phase will be triggered to collect runtime data, like local storage data, function innovation details and dynamic HTML content. 
We will detail the \textit{page navigation} and \textit{runtime data collection} in detail in the following.

\textbf{Page Navigation.} 
Specifically, two navigation routes are designed, according to the lifecycle of a wallet, as shown in Figure~\ref{fig:wallet activity}. 
One regards the wallet as a newly created one and the path is mainly composed of wallet creation and mnemonic display. Another path assumes the wallet is imported, i.e., including mnemonic import, password setting, wallet unlock and wallet backup. 
Specific operations for each page are set through the Selenium framework~\cite{selenium}, a famous framework for automating web browsers. Except for the ``mnemonic display page" and ``mnemonic backup page", which are respectively the ends of two navigation paths, we customize the operation scripts for the remaining 11 pages. 
Upon arriving at a particular page, the corresponding script is triggered. 
Although pages with different semantics will be navigated in different manners, the script is designed to follow a general strategy. 
Firstly, the script locates the interactive elements with their labels, such as buttons, checkboxes and input fields, etc., which helps {\tool} gain an overall understanding of the current page and facilitate further user action simulations. 
Secondly, it manages pop-ups and checkboxes by clicking on prompts like ``continue'' and this reflects real-world scenarios where users frequently encounter and interact with such elements for confirmations or agreements. 
Thirdly, it comes to the input fields of the current page. The script systematically populates them with predefined credentials, such as usernames and passwords. This step is particularly important for testing specific pages like the password setting or mnemonic import pages, as it allows {\tool} to test these functionalities as expected and advance the testing path accordingly; 
in cases where labels correspond to fields not covered by our predefined credentials (such as wallet nicknames), the script will generate and input random strings to help complete the functionality of the page. 
At last, the script seeks to navigate to the next pages by engaging elements like ``Next'' or ``OK'', facilitating {\tool} in successfully completing the predefined navigation route. 

\textbf{Runtime Data Collection.}
During these processes, the instrumented code is triggered to collect runtime data and sensitive data generated during the interaction, like wallet passwords and mnemonics. 
Besides, during the page navigation, {\tool} takes advantage of the periodic call function in JavaScript to continuously monitor the changes of the current HTML page and local storage (including LocalStorage, SessionStorage, IndexedDB and local session files).
The monitor is performed once a second whenever modifications occur. 

\begin{figure}[htbp] 
\centering
\includegraphics[width=0.8\linewidth]{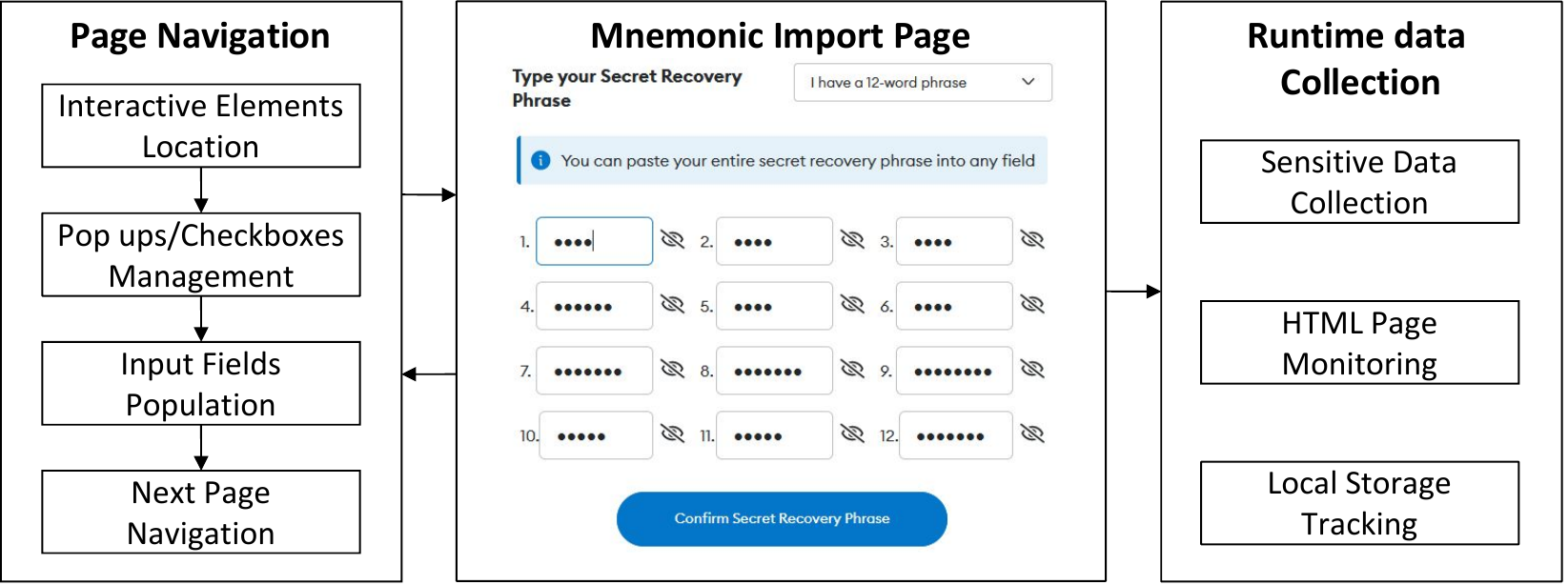} 
\caption{\hl{The page navigation and runtime data collection on a mnemonic import page.}} 
\label{fig:pagenavi} 
\vspace{-0.2in}
\end{figure} 
 
\hl{Take the mnemonic import page as an example to illustrate how the page navigation and runtime data collection are conducted, the process of which is shown in} Figure~\ref{fig:pagenavi}. 
\hl{On this page, the major interactive element the WalletRadar locate will be a certain number (12, 15, or 24) of input boxes for entering mnemonics (12 boxes in the example)}, or a single text box to receive all the mnemonics separated by spaces; next, after handling pop-up windows and checkboxes, WalletRadar fills in the input boxes or the text box with the mnemonic words generated during the previous wallet creation; for other non-sensitive information (such as the wallet name), random strings are generated and entered. \hl{The import process is completed by pressing the confirm button (blue button in the exmaple)}. Due to code instrumentation during the static analyzing phase, {\tool} will capture the HTML code of the front-end page during mnemonic import and log the local storage data after the mnemonic input. After the wallet import, the wallet will navigate to the home page and then progress to test the wallet unlock page.

\subsection{Vulnerability Detecting} 
\label{sec:approach:vul} 
In the vulnerability detecting phase, {\tool} takes advantage of information collected from the previous static analyzing and dynamic analyzing phases to identify potential vulnerabilities hidden in the given wallets. 
Specifically, {\tool} integrates six rule-based detectors, corresponding to the six vulnerabilities mentioned in \S\ref{sec:taxonomy}. We argue that {\tool} can be easily extended by implementing detectors based on the collected information. 
The detecting strategies are detailed in the following. 

\textbf{Clickjacking.} 
As depicted in \S\ref{sec:clickjacking}, clickjacking in browser-based wallets is related to the ``web\_accessi\-ble\_resource'' item in the ``manifest.json'' configuration file. 
If sensitive HTML pages, such as the home page and transaction page, are present in the configuration item according to the static analysis, it can be concluded that these web pages can be accessed by external websites (including phishing sites) and the wallet is exploitable to this vulnerability. 

\textbf{XSS Vulnerability.} 
According to the guidance provided by PortSwigger~\cite{domxss}, during the static analyzing phase, {\tool} traces back from DOM-manipulating functions to the data source. 
If the data source function at the end of this trace is found in the valuable function database and is identified as being susceptible to external modification, it signifies a possible DOM-based XSS vulnerability. This scenario highlights a risk where the content within the wallet’s web pages could be externally altered or compromised, making the wallet vulnerable to XSS attacks. 

\textbf{Defective Password Policy.} 
In the dynamic analyzing phase, when encountering a password setting page, the {\tool} will try to test a set of passwords on the page, which start from the weakest one (e.g., ``123'') to the relatively strong one (e.g. ``Weasdxz@a142''). It will then record the weakest password that finally passes the password setting of one wallet. According to the CheckPoint report~\cite{checkpoint}, if the password is composed of only 6 or fewer digits, the strength of this password is considered insufficient for supporting the security of the browser-based wallet. 

\textbf{Redundant Storage.} 
During the dynamic analyzing phase, {\tool} continuously records all intermediate data. 
Thus, this detector compares the data stored in the local storage and the intermediate data generated during the decryption. 
If the intermediate data can be matched to part of the local storage data, the wallet may leak sensitive information, i.e., it is vulnerable to redundant storage vulnerability. 

\textbf{Demonic Vulnerability.}
This detector focuses on the textual elements (e.g., ``textarea'' tags) in front-end HTML pages. 
To detect demonic vulnerabilities, the detector scans for specific textual elements on HTML pages during sensitive operations, particularly focusing on those that hold plaintext mnemonics or private keys. If these elements are found and corresponding plaintext data is also present in the browser's local storage, the wallet is flagged for demonic vulnerability. 

\textbf{Defective cryptography.}
If the detector finds that the wallet applies PBKDF2 with less than 10K rounds or uses inappropriate methods like AES-CBC mode, the wallet is believed to have a defective cryptography vulnerability.

\section{Evaluation} 
\label{sec:vul} 
In this section, we conduct a comprehensive evaluation on {\tool} to characterize the vulnerabilities in browser-based wallets in the wild. 

\subsection{Research Questions \& Experimental Setting} 
In this paper, we are interested in the following questions:

\begin{itemize}[leftmargin=9mm]
    \item[\textbf{\hl{RQ1}}] \textbf{\hl{Is WalletRadar efficient and effective in detecting these vulnerabilities?}}
    \item[\textbf{\hl{RQ2}}] \textbf{\hl{What are the characteristics of vulnerable wallets in the wild?}}
\end{itemize}

To answer RQ1, since there are no existing datasets for browser-based wallets, we collect a dataset from the Chrome Web Store~\cite{webstore}. We manually inspect these samples to build a reliable benchmark to evaluate the efficiency and effectiveness of {\tool}.
To answer RQ2, based on the detection results, we analyze the characteristics of these vulnerabilities, evaluate their impacts and track the developers' responses and remediation efforts.

\textbf{Experimental Setting.} We implement {\tool} based on Python3. In specific, {\tool} utilizes js-beautify~\cite{jsbeauty} for code formatting, Esprima~\cite{esprima} for AST parsing of JavaScript, and the Selenium framework~\cite{selenium} for invoking automated scripts. Other analytical components, including automated runtime scripts, valuable function databases, and detection rules, are all designed independently. 
The following experiment is performed on a laptop with the Intel Core i7-12700H@2.3GHz Processor and 16G RAM. The Selenium framework is operated on a Chrome web browser (version:102.0.5005.189) to test browser-based wallets dynamically.

\textbf{Dataset Collection.} 
We collect browser-based samples from the Chrome Web Store, one of the most widely adopted and well-known platforms for browser extensions. 
There are 618 results when searching ``blockchain wallet''. To produce an effective result, we conduct a filtering process based on some criteria.
First, as our subject is the browser-based non-custodial wallet, other types of wallets are not considered. 
Second, these wallets need to be popular. Thus, we keep the wallet with more than 3K users. 
To cover as many use cases as possible, we consider wallets that support either a single or multiple blockchains. 
At last, we have collected 120 samples in total. 
However, we find some of them are not fully functional, such as failure at wallet creation or mnemonics import. 
Thus, based on user comments on the web store, we removed the samples with bad reputations. 
Consequently, 96 samples are regarded as candidates, corresponding to multiple blockchains like Bitcoin, Ethereum, and Solana. \hl{Considering the limited number of samples and the absence of established vulnerability ground truth in our dataset, we opt for a comprehensive manual analysis of each sample. To build a trustworthy benchmark, the manual labeling is conducted independently by two experienced authors who are familiar with typical vulnerability signatures in browser-based wallets. By doing so, we precisely identify and classify relevant vulnerabilities, thereby creating an accurate benchmark dataset to guide future analyses and comparisons.}

\textbf{Dataset Overview.}
In summary, these samples have been downloaded at least 23 million times in total, accounting for about 97\% of total downloads in the search results of ``blockchain wallet'' on the Chrome Web Store. 
This indicates that a browser-based non-custodial wallet is the major choice of blockchain users. The top 10 browser-based wallets according to download times are shown in Table~\ref{tab:top10wallet}. 
As we can see, they have at least 500K users and support for more than 10 blockchains in total, highlighting their widespread popularity and versatility in catering to diverse blockchain platforms and user needs.

\begin{table}[t]
\caption{The top 10 browser-based wallets that most users download.}
\centering
\label{tab:top10wallet}
\begin{tabular}{@{}llll@{}}
\toprule
Name     & Version & Users & \begin{tabular}[c]{@{}l@{}}Supported \\ Blockchains\end{tabular} \\ \midrule
Metamask & 10.14.0 & 10M+  & Ethereum, Polygon, ...                                        \\
Phantom &  22.9.6     &  2M+     &  Solana, Ethereum, ...                                                             \\
Ronin Wallet       &  1.23.1     &   1M+    &    Ronin                                                           \\
Binance wallet         &  2.13.7     & 1M+      &  BNB Chain, Ethereum, ...                                                             \\
Coinbase         &   2.30.2    &  1M+     &   Ethereum, Avalanche, ...                                                            \\
 Keplr        &  0.11.1     &  900K+     &    Osmosis, Mars, ...                                                           \\
 Station        &  3.1.0     & 600K+      &  Terra                                                             \\
Argent X         &  5.2.0     &   600K+    &   Ethereum                                                            \\
TronLink         &  3.26.9     &  500K+     &   Tron                                                            \\
Martian Wallet         & 0.2.2      &  500K+     &  Sui, Aptos                                                             \\ \bottomrule
\end{tabular}
\end{table}

\subsection{RQ1: Efficiency \& Effectiveness} 
In this section, we evaluate the efficiency and effectiveness of {\tool} on the collected 96 samples when identifying vulnerabilities. 

\subsubsection{Automation Efficiency Test} 
We performed automated tests on these 96 samples using {\tool}. Among them, 9 samples only completed the static analyzing phase because of their unique attributes during creating wallets, importing wallets, etc., preventing the full completion of automated dynamic operations. \hl{For example, one wallet may require a long press to export mnemonic phrases while another directs users to an external website during page navigation, both scenarios disrupting typical test workflows}. However, the remaining 87 samples still successfully completed all the automated analysis processes, achieving a 90.6\% automation completion rate. We also manually intervened on the 9 samples that failed to complete the dynamic analyzing phase to ensure they underwent full vulnerability detection. In the automation testing process, the execution time for each sample was approximately 8 minutes, with around 5 minutes for the static analyzing phase and about 3 minutes for the dynamic analyzing phase. This suggests that {\tool}'s automation testing is relatively swift, owing to its clear execution paths and operations. 

\begin{table}[t]
\centering
\caption{The detection results of 96 samples.} 
\label{tab:detection results}
\begin{tabular}{cccc}
\toprule
Category &  
  \begin{tabular}[c]{@{}c@{}}\# of \\ Vulnerabilities\end{tabular} &
  \begin{tabular}[c]{@{}c@{}}False \\ Positives\end{tabular} &
  \begin{tabular}[c]{@{}c@{}}False \\ Negatives\end{tabular} \\ \midrule 
Demonic Vulnerability & 55 & 0 & 0 \\ 
Defective Password Policy & 20 & 0 & 0\\ 
Redundant Storage & 18 & 0 & 0 \\ 
Clickjacking & 13 & 0 & 0 \\ 
Defective Cryptography & 2 & 0 & 4 \\ 
XSS & 2 & 0 & 2 \\ 
\midrule
Total & 110 & 0 & 6 \\ \bottomrule 
\end{tabular}
\end{table} 

\subsubsection{Evaluation of Detection Results} 
\label{sec:detection_overview}

The detection results on 96 samples are shown in Table~\ref{tab:detection results}. 
As we can see, a total of 110 vulnerabilities in 70 samples (73\% of all tested samples) are identified by {\tool}. 
The demonic vulnerability becomes the most serious problem among these wallet extensions, which accounts for 57\% of all tested samples. \hl{This may result from the situation that many wallet extensions forked the early versions of the famous wallet Metamask, which previously contained a demonic vulnerability.} The other three slightly less serious issues are defective password policy (20), redundant storage (18), and clickjacking (13). This suggests that both wallet-specific vulnerabilities and traditional web vulnerabilities are prevalent on wallet extensions. Among them, 53\% (37) of the wallets have one vulnerability and the wallet with the most vulnerabilities has 5 distinct issues except defective cryptography. 

\hl{As for the false alarms, we define that a false positive happens when WalletRadar flags a non-existent vulnerability, and a false negative occurs when it misses an actual vulnerability. To ensure a precise evaluation of WalletRadar's detection capabilities, the evaluation process involves a detailed review of the target wallets' code and functionalities to identify known vulnerability patterns and compare them against the tool's findings. After the evaluation, we discover that there are no false positives for all six vulnerabilities.} We speculate this is due to the characteristics of these vulnerabilities and the efficiency of {\tool}. 
Leveraging the comprehensive valuable function database at the outset enhances the accuracy of our detections, complemented by the use of specific static rules that precisely identify unique vulnerability patterns. Additionally, our accurate dynamic testing methods distinguish between true vulnerabilities and normal behaviors. 
Moreover, we can easily observe that {\tool} cannot fully detect XSS and defective cryptography vulnerabilities. 
After manual verification, we discover that the four defective cryptography cases include two ``instances of insecure AES usage" and two instances of ``insufficient iterations". 
The cryptographic parameters of these four samples are generated dynamically within closures or local scopes, instead of being hardcoded, thus they evade the static analysis. The dynamic instrumentation, designed to capture parameters at runtime, also failed in these cases because the parameters were generated within specific execution contexts or transient states that were not active or accessible during the instrumentation phase, preventing their capture and analysis. 
As for the two false negatives in the XSS vulnerability, we find that the data sources and sink points are spread across different files. Therefore, the AST-based analysis method struggles with such inter-file detection, leading to these oversights.

\begin{framed}
 \noindent \textbf{RQ1 Answer:} Our experiments, conducted on 96 widely-used cryptocurrency wallets, demonstrate that our tool, {\tool}, can automatically complete the detection process on over 90\% of these wallets, with the capacity to cover all wallets when supplemented by manual intervention. Furthermore, {\tool} exhibits high accuracy in our collected dataset, characterized by the absence of false positives and a minimal incidence of false negatives. In summary, {\tool} proves to be both efficient and effective in identifying vulnerabilities.
 
\end{framed}

\subsection{RQ2: Characteristics of Vulnerable Wallets in the Wild} 
\label{sec:vuldetail}
In this section, we first characterize how these vulnerabilities spread in the wild. Then, we evaluate the impact directly brought from them.

\subsubsection{Studies on Real-world Vulnerabilities} 
\label{sec:vulncase} 

\begin{figure}[t]
\centering
\includegraphics[width=0.6\textwidth]{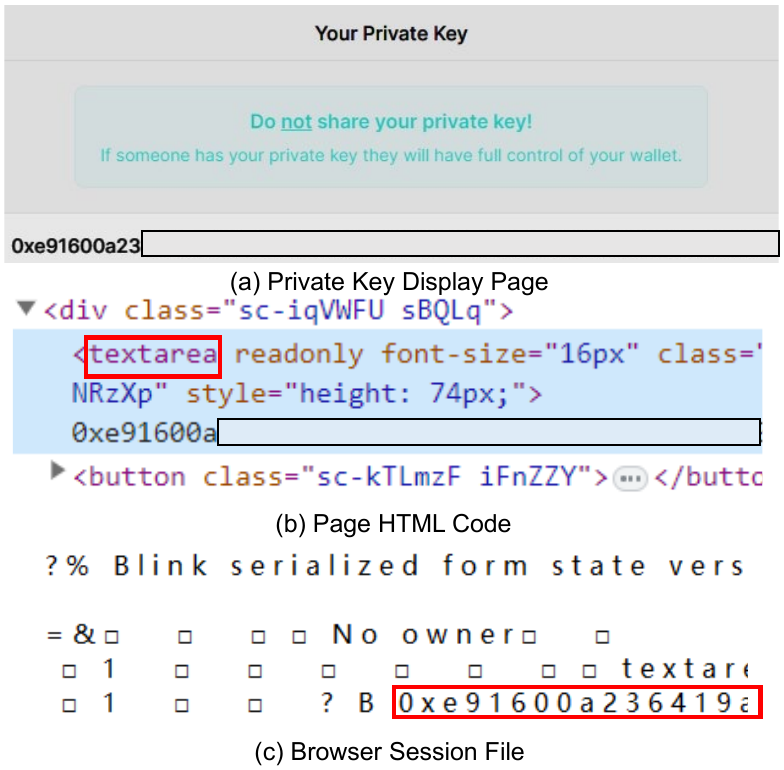}
\caption{The demonic vulnerability in P* Wallet.}
\vspace{-0.1in}
\label{fig:demonic_example}
\end{figure}

\textbf{Demonic Vulnerability.} 
During our evaluation, it was found that 55 wallets use textual HTML tags to store sensitive information when backing up, importing or displaying mnemonics and private keys. 
Take P* Wallet\footnote{Due to ethical considerations, we have anonymized the names of the wallets, displaying only their initial capital letters.} in Figure~\ref{fig:demonic_example} as an example, when a user displays the private key, the wallet uses textual tags to store the private key, and this data will be cached in the local storage, posing a threat of sensitive data leakage. The way to fix this vulnerability is to use 12-24 input boxes with ``input'' tags and a ``password'' attribution in an HTML page to accommodate the user's mnemonic or private key. When this implementation is adopted, the user needs to perform multiple inputs, which may be cumbersome and require developers to invest more effort in mnemonic input optimization. 

\textbf{Defective Password Policy.} 
Among the 20 samples with password policy flaws, 1 sample requires a minimum of 4 digits, 9 samples do not require password complexity, and 10 samples require a minimum of 6 digits. It is worth mentioning that among the 96 samples, only 29 samples have the minimum password complexity requirement of mixed letters and numbers. In real-world scenarios, there is a high likelihood of weak passwords in browser-based cryptocurrency wallets, resulting in more accessible brute-force cracking and endangering the security of users' wallets. For building a secure cryptocurrency wallet, it is wise to require a minimum of 8-digit password with a mixture of numbers and letters.

\textbf{Redundant Storage.} 
All cases of this vulnerability are found in the process of unlocking the cryptocurrency wallet, and the evaluation results show that this vulnerability is severe in actual situations. 18 samples demonstrated varied manifestations of this vulnerability: 

\one Embedding the key required by the decryption function within the code. 

\two Storing all raw materials needed for the decryption function in local storage. 

\three Using symmetric encryption based on a timestamp to save the user's password in local storage, where the timestamp is stored in plaintext. 

\four Encrypting and storing the user's password in local storage using a fixed key in symmetric encryption, with the fixed key embedded within the code. 

\five Storing the user's plaintext password directly in local storage. 

\six Storing the initial hashing result of the user's password in local storage (without using it for decryption). 

\begin{figure}[t]
\centering
\includegraphics[width=0.6\textwidth]{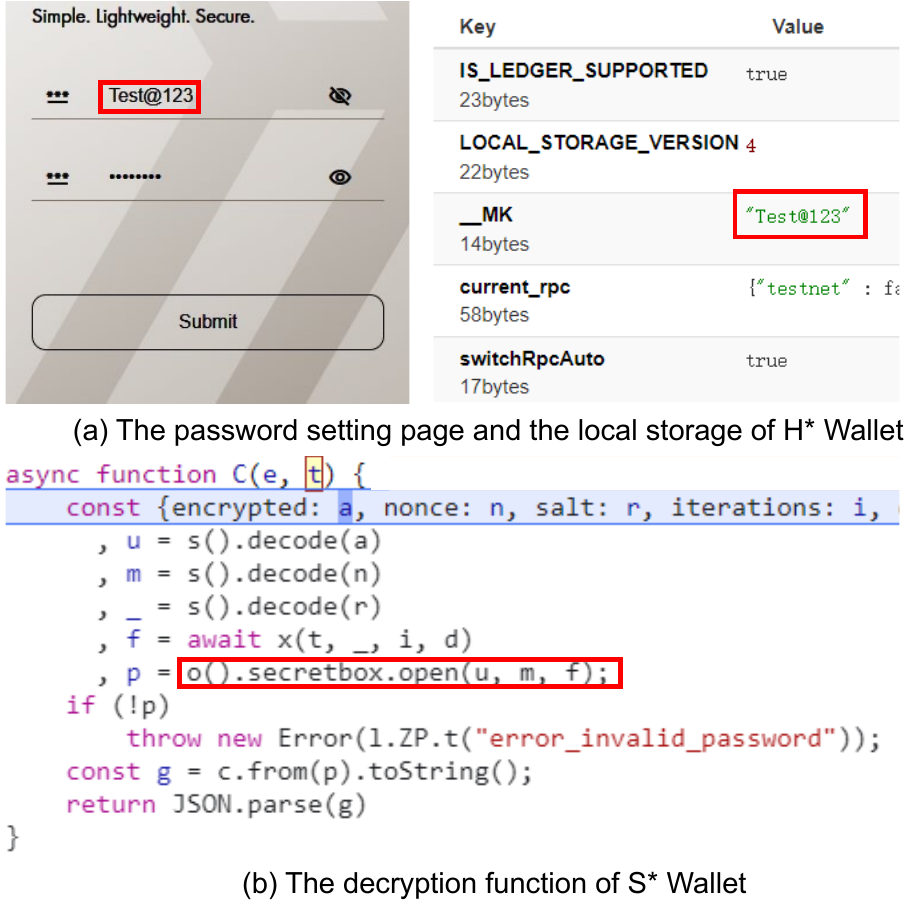}
\caption{The redundant storage vulnerability in two wallets.}
\vspace{-0.1in}
\label{fig:redundant_example}
\end{figure}

The first five cases can directly result in users losing control of their wallets if an attacker gains access to the stored content. In the sixth case, although the wallet stores the initial hash result (e.g., SHA-512) of the user's password in local storage, it does not use the hash for decryption. Instead, when the user enters their password, the wallet performs the same hashing process and compares the two hash results for password verification. Although this practice cannot directly lead to the decryption of the user's wallet, it enables an attacker to extract the hash and use it for brute force cracking, posing a severe threat to the user's wallet. Figure~\ref{fig:redundant_example}(a) is an example where the H* Wallet directly saves the user's password with plaintext in the local storage, making the wallet vulnerable to attacks. For S* Wallet, although it saves the processed password in the local storage instead of plaintext ones, the data can be directly used in its decryption function shown in Figure~\ref{fig:redundant_example}(b) to get the mnemonics, which is also an insecure practice. 

All six cases are related to poor practices of wallet developers when they deal with sensitive data storage and the fifth case is the worst one, which fully exposes the credentials. It is recommended to keep the decryption process in real-time and not save any sensitive data in local storage just for ease of use.

\textbf{Clickjacking.} 
13 wallet samples were found with clickjacking vulnerabilities, which indicates that such vulnerabilities in the actual situation are also very prominent. We have identified two situations where these cryptocurrency wallets have such vulnerabilities: the introduction of the ``wallet home page" and the ``security reminder page" in the ``web\uline{ }accessible\uline{ }resources” configuration. Some wallets directly introduce the ``wallet home page" in configuration, and this kind of implementation makes the home pages of these wallets vulnerable to hijacking attacks, causing users to unwittingly download malware, visit malicious web pages or provide sensitive information. Besides, while some wallets also provide users with some security detection services (phishing detection, etc.) and click-jumping functions, they inadvertently introduce this type of vulnerability as a consequence. Figure~\ref{fig:clickjacking_example} shows a basic example that exploits the vulnerability. The code can be embedded in a phishing website to overlay the wallet's phishing warning page, inducing users to jump to the wallet home page and perform sensitive operations.

The original purpose of introducing files in the ``web\uline{ }accessible\uline{ }resources” configuration is to expose the resources (such as images) for external web pages to access. However, when introducing HTML files, other external pages can directly access the web pages and there is a possibility that these HTML pages are embedded in a phishing website, resulting in clickjacking. The wallet developers need to avoid adding HTML files to this configuration or ensure that the HTML files added do not contain key wallet functions such as importing wallets, sending transactions, etc. 

\begin{figure}[t] 
\centering
\begin{lstlisting}[style=styleHTML]
<iframe src="chrome-extension://{extension id}/phishing.html?href=chrome-extension://{extension id}/wallet.html" width="100%" height="100%">
</iframe>
\end{lstlisting}
\caption{The code implementation of the clickjacking in S* Wallet.}
\label{fig:clickjacking_example}
\end{figure}

\begin{figure}[t]
\centering
\begin{lstlisting}[style=styleJS]
function a(e, t) {
    var o = n.utf8ToBuffer(e)
      , i = n.base64ToBuffer(t);
    return r.crypto.subtle.importKey("raw", o, {
        name: "PBKDF2"
    }, !1, ["deriveBits", "deriveKey"]).then((function(e) {
        // Generate the key for decryption iteratively
        return r.crypto.subtle.deriveKey({
            name: "PBKDF2",
            salt: i,
            iterations: 5e3,
            hash: "SHA-256"
        }, e, {
            name: "AES-GCM",
            length: 256
        }, !1, ["encrypt", "decrypt"])
    }
    ))
}
\end{lstlisting}
\caption{The code snippet that has a defective cryptography vulnerability.}
\label{fig:generate key code}
\end{figure} 

\textbf{Defective Cryptography.} 
In cryptographic practices, it is generally advised to employ algorithms like argon2, scrypt, and PBKDF2 with a higher number of iterations to derive keys based on users' passwords. In our study, while most wallets adhere to the recommended $10,000$ iterations for PBKDF2, three samples fall below this standard, using fewer than $5,000$ iterations, one of which is depicted in Figure~\ref{fig:generate key code}. Only five samples demonstrated exceptional security with $310,000$ iterations. This suggests that most wallets generally meet cryptographic standards but often do not implement the highest level of security practices. Combined with inadequate password policies, this shortfall in implementing the highest security standards increases the risk of brute-force attacks. 

Besides, the majority of wallets employ AES for symmetric encryption and decryption. We discovered only three instances that used the less secure CBC mode, while the rest followed best practices by using GCM or CTR mode. 

\begin{figure}[t]
\centering
\begin{lstlisting}[style=styleJS]
window.onload = function() {
    if ("/phishing.html" === window.location.pathname) {
        // Extract the "hostname" parameter and assign to the variable "e"
        const {hostname: e} = h();
        // Write the parameter value directly into the HTML page
        document.getElementById("esdbLink").innerHTML = '<b>To read more about this scam, navigate to: <a href="https://etherscamdb.info/domain/' + e + '"> https://etherscamdb.info/domain/' + e + "</a></b>"
    }
}

function h() {
    // Parse the hash in the current URL
    const e = window.location.hash.substring(1);
    return o.parse(e)
}
\end{lstlisting}
\caption{A code snippet that has an XSS vulnerability.} 
\label{fig:KardiaChain Wallet phishing code}
\end{figure}

\textbf{XSS Vulnerability.} 
Among the 96 samples, only 4 samples were detected with DOM-based XSS vulnerabilities (one of whose code is shown in Figure~\ref{fig:KardiaChain Wallet phishing code}). These vulnerabilities occurred in HTML pages associated with clickjacking and were related to the security reminder function provided by the wallet, indicating that exposed HTML pages (introduced in the ``web\uline{ }accessible\uline{ }resources" configuration) are more likely to have XSS vulnerabilities in the browser-based cryptocurrency wallets. Notably, although the code of these samples suggests that they are vulnerable to XSS attacks, further manual inspection reveals that their adoption of the Content Security Policy (CSP), the added layer of security, mitigates the impact of this kind of vulnerability.

\subsubsection{Impact} 
Based on the analysis result, we find 70 vulnerable wallets with at least 9.2 million downloads. Among these wallets, the most popular one achieves more than a million downloads and 23\% of these wallets have more than 100K users. If these wallets are attacked, a large number of users will be exposed to the risk of information leakage and even financial losses. Thus, when we finished the analysis of these 70 examples in February 2023, we reproduced the vulnerabilities to confirm their existence and attempted to get in touch with the developers to report these issues. Finally, we got confirmations of 10 vulnerabilities from the developers of 8 wallets with over \$2000 bounties. By study time, we revisit the wallets with vulnerabilities and check whether these vulnerabilities have been fixed. \hl{The outcomes of these checks are detailed in} Table~\ref{tab:fix results}\hl{, presenting an overview of the vulnerabilities' current status.}

\begin{table}[h]
\centering
\small
\caption{\hl{The current status of identified vulnerabilities in browser-based wallets.}}
\label{tab:fix results}
\begin{tabular}{@{}cccc@{}}
\toprule
Vulnerabilities & \begin{tabular}[c]{@{}c@{}}\# of \\ Vulnerabilities\end{tabular} & Fixed & Confirmed \\ \midrule
Demonic Vulnerability     & 55  & 18 & 6  \\
Defective Password Policy & 20  & 6  & 3  \\
Redundant Storage         & 18  & 3  & 1  \\
Clickjacking              & 13  & 6  & 0  \\
Defective Cryptography    & 6   & 3  & 0  \\
XSS                       & 4   & 0  & 0  \\ \midrule
Total                     & 116 & 36 & 10 \\ \bottomrule
\end{tabular}
\end{table}

\textbf{Active Wallets without Fixing the Issues.} We refer to the wallets that have been updated in the last six months as ``active wallets''. For 29 active wallets, our manual inspection suggests that their developers are mainly involved in updating features or fixing bugs in functionality, such as adding support for DApps, launching promotion campaigns for the new tokens, or integrating with cryptocurrency exchanges, etc. Besides, Some developers believe it is the users' responsibility to keep the wallet safe and refuse to fix weaknesses related to password policy and cryptography. 
Although wallet users control their keys and are primarily responsible for keeping them safe, expecting all users to have good password management habits is unrealistic. Thus, wallet developers should also pay as much attention as possible to the security of the wallets in order to set the stage for wallet users and mitigate the potential risk to users. 

We also find that some wallets proclaimed to have fixed demonic vulnerabilities but only fixed those mentioned in Metamask blog~\cite{metamask_demonic}, overlooking other functions with demonic vulnerability. Figure~\ref{fig:demonic_example} presented in \S\ref{sec:vulncase} shows the example of P* Wallet, which leaves the private key display page unfixed. This suggests that some developers lack comprehensive knowledge of vulnerability-related information, resulting in incomplete vulnerability fixes. 

\textbf{Active Wallets with Issues Fully/Partly Fixed.} Among 26 active wallets, developers of 8 wallets have confirmed 10 vulnerabilities of their wallets to us and all of them have fixed the issues by study time. \hl{5 wallets fully fixed 5 issues before our reports and 10 wallets addressed 14 vulnerabilities post-report silently, with 4 of the fixes occurring before our contact.} Among the remaining 3 wallets, none have fully addressed their vulnerabilities. Two wallets (including the wallet with the most vulnerabilities mentioned in \S\ref{sec:detection_overview}) have only resolved the demonic vulnerability silently after our reports, leaving other vulnerabilities unfixed. \hl{The developers' choice to fix severe vulnerabilities like the demonic vulnerability first, both in terms of proportion and sequence, may suggest they prioritize the most critical issues, intending to address less severe vulnerabilities in future updates.}

Besides, one wallet fixes the clickjacking vulnerabilities before our report while still leaving its 6-digit password policy unchanged after the report. We speculate this is because the developers may have been concerned that fixing the password policy would create backward compatibility issues, or simply think that the current password strength is sufficient.

\textbf{Inactive Wallets.} For the remaining 15 wallets that are not updated or removed from the Chrome web store, we find some of these wallet developers have moved to new projects after constructing a basic feature of their wallet extensions while some just stopped the maintenance. However, as non-custodial wallets can operate without central services, the users who still use these wallets may suffer from unsuspecting attacks. Considering at least 314K users that have ever used these wallets, there may be a number of users who are at risk of being attacked. 

In summary, our vulnerability disclosures result in updates to 20 wallets by their developers, which include developers of 8 wallets who confirmed and fixed the issues and developers of 12 wallets who silently patched the vulnerabilities, fixing a total of 26 vulnerabilities. This accounts for 22.4\% of the vulnerabilities we identified, protecting their thousands of users from potential attacks. However, it is worth noting that many developers still do not place enough emphasis on the security development process of their wallets, thereby potentially compromising the security of their user base. 

\begin{framed}
\noindent \textbf{RQ2 Answer:} \hl{The case studies of 116 real-world vulnerabilities reveal a widespread prevalence of security issues in browser-based wallets. Notably, the demonic vulnerability becomes most critical and it is often overlooked by developers regarding its potential occurrence. Other vulnerabilities also shed light on a mix of wallet-specific and general web vulnerabilities within these browser extensions. This underscores a potential deficiency in secure development practices among their developers.} The conducted impact analysis reveals that these vulnerabilities pose a significant risk to approximately 9.2 million users, with threats ranging from information leakage to financial loss. In response to our disclosure of these vulnerabilities, there has been a rectification of 26 vulnerabilities across 20 different wallets, thereby mitigating their possible adverse effects. 
\end{framed}

\section{Discussion}
\label{sec:discussion}

\subsection{Implications} 

The work on characterizing and detecting security issues in browser-based cryptocurrency wallets is essential for stakeholders in the community. 

For developers and wallet users, the taxonomy of security issues in browser-based cryptocurrency wallets assists them in understanding potential vulnerabilities, offering a valuable reference throughout the application's security lifecycle. \hl{Besides, despite the diversity in programming languages and platforms, other kinds of blockchain wallets and blockchain applications often share common operational workflows or mechanisms. For example, many desktop wallets also use the same key management techniques (e.g., using PBKDF2 for key generation) as browser-based wallets. Thus, certain components of WalletRadar, including automated testing modules and identified vulnerability patterns, might be adapted for use in other applications with some modifications.} 

The evaluation of our proposed detection framework demonstrates the effectiveness of our proposed approach while suggesting that a large number of current browser-based wallets suffer from various security vulnerabilities. Besides, although \tool{} is primed to detect vulnerabilities identified in this paper, it also retains high extensibility. For instance, the vulnerable function database and the rules for the vulnerability detector can be added to or optimized, facilitating targeted detection of subsequent vulnerabilities related to browser-based wallets. Combining this approach with web application audit methods, as described in this paper, can help reduce vulnerabilities during wallet development and lower the risk of financial loss for users. 

Further, our impact analysis of these vulnerabilities suggests that millions of users may be vulnerable to attacks related to these vulnerabilities, which could result in substantial financial losses. Moreover, our subsequent analysis of these wallets over time after reporting these vulnerabilities shows that only a small number of developers fully fix these vulnerabilities, while some serious vulnerabilities such as the demonic vulnerability are even mistakenly believed to be fixed due to developers' superficial knowledge of them. Developers should take wallet security issues more seriously and gain a better understanding of these vulnerabilities to effectively fix them and fully mitigate their impact. 

\hl{We recognize the ethical aspects of detecting vulnerabilities in browser-based wallets and maintain confidentiality by anonymizing application names. Our responsible disclosure ensures developers have time to fix issues before we make them public, and we provide repair strategies for all vulnerabilities. Our goal is to improve the security of the blockchain ecosystem, making it safer for users.}

\subsection{Limitations} 
Firstly, we initially established a taxonomy of browser-based cryptocurrency wallet security issues based on information collection from various channels (as described in \S\ref{sec:taxonomy}). While we have done our best to refine the taxonomy, it may still be incomplete. Nevertheless, the utility and effectiveness of our proposed taxonomy were confirmed through evaluation, revealing numerous security issues faced by current browser-based cryptocurrency wallets. 

\hl{The current keyword-based semantic approach for identifying key wallet pages in dynamic analysis might fall short in more complex semantic contexts. More advanced techniques such as optical character recognition (OCR) and large language models (LLMs) could enhance page traversal capabilities by understanding and interpreting complex page content more effectively. These techniques promise to refine the automation process, potentially increasing the coverage beyond the current 90\% and reducing the need for manual intervention.}

Besides, as most browser-based wallets are non-custodial wallets and their users' addresses are not publicly known, it is hard for us to track whether these wallets are actually under attack or not on the blockchain. Nevertheless, our experiments and the following investigations indicate that these vulnerabilities are still not fixed and could have the potential to cause financial loss to a large number of users.

\section{Related Work} 

\subsection{Web Application Analysis} 
Browser extensions for cryptocurrency wallets are essentially web applications, leading us into a broader discussion on web application analysis in software engineering. Since it has been in development for many years, there has been a lot of work dedicated to studying web applications and developing related analysis tools. Some efforts focus on automatic web application testing~\cite{nguyen2019exploring,zheng2021automatic,lin2023automated}. For example, Zheng et al.~\cite{zheng2021automatic} presents an end-to-end automated web testing framework. Using curiosity-driven reinforcement learning, it efficiently generates high-quality action sequences for web application testing. 
Another major and more relevant research direction to this work is developing automated detection frameworks to detect security issues in web applications like XSS~\cite{pan2017detecting,eriksson2022hardening,pazos2023xsnare} or privacy breach related vulnerabilities~\cite{zhao2015automatic,chen2018mystique,kariryaa2021understanding}. For example, Pan et al.~\cite{pan2017detecting} propose a detecting framework employing hybrid analysis combined with lightweight static analysis consisting of a text filter and an abstract syntax tree parser and dynamic symbolic execution to detect DOM-based XSS vulnerabilities. To detect privacy leaks, Chen et al.~\cite{chen2018mystique} propose a hybrid taint analysis technique that leverages both dynamic taint tracking and static analysis by using information gathered from static data flow and control-flow dependency analysis to propagate taint at runtime. Besides, Kariryaa et al.~\cite{kariryaa2021understanding} find that users have limited technical knowledge about browser extensions' capabilities and they are only focused on the features these extensions provide. 

The above work provides many insights into this paper for the vulnerability taxonomy and the detection methods of browser extension vulnerabilities. Based on these static and dynamic analysis methods, this paper customizes and optimizes the {\tool} to suit the specific needs of browser-based cryptocurrency wallets, addressing their distinct vulnerabilities and operational processes. 

\subsection{Cryptocurrency Wallet Analysis} 
Current research on cryptocurrency wallet mainly focuses on mobile application wallets~\cite{sai2019privacy,he2020security,li2020android,praitheeshan2020security,hu2021security,uddin2021horus,wang2022characterizing}. For example, Sai et al.~\cite{sai2019privacy} first used static code analysis and network data analysis to evaluate the security issues of Android-based cryptocurrency wallets, and found that the security of popular cryptocurrency wallet apps is not significantly worse than banking apps, but they lack privacy protections. Li et al.~\cite{li2020android} assessed the security issues of cryptocurrency wallet apps and presented a comprehensive attack surface on them. He et al.~\cite{he2020security} carried out related attack experiments on the premise that the attacker can access the user's mobile device with high privileges. 

In addition to researchers focusing on mobile wallets, Praitheeshan et al.~\cite{praitheeshan2020security} conducted a security assessment of smart contract-based on-chain Ethereum wallets. With the help of automated scanning tools, they conducted a security analysis on the wallets on the chain using smart contracts and gave a classification of security issues. Guri M et al.~\cite{guri2018beatcoin} conducted a security analysis on air-gapped cryptocurrency wallets (i.e., wallets isolated from the Internet) and proved that attackers may still attack isolated offline wallets to steal private keys through various exfiltration techniques. 

There is a lack of systematic security analysis of browser-based cryptocurrency wallets. Considering the large user base of browser-based wallets, it is necessary to get an understanding of them and develop an automated security assessment tool to help developers deal with potential vulnerabilities during the development process. In this work, we take the first step to characterize and detect vulnerabilities in browser-based wallets.

\section{Conclusion} 
This paper presents the first systematic assessment of vulnerabilities in browser-based cryptocurrency wallets. We propose {\tool}, an automated detection framework leveraging a hybrid of static and dynamic analysis to efficiently identify vulnerabilities. The experiments on popular browser-based wallets demonstrate that {\tool} operates automatically on the majority of these wallets with high accuracy. The evaluation results have also uncovered widespread security issues, highlighting a concerning lack of awareness among many developers regarding these vulnerabilities. Unfortunately, only a few developers have thoroughly addressed these security flaws.

\balance
\bibliography{sigproc} 

\section*{Declarations} 

\noindent \textbf{Ethical Approval} Not applicable since there are no human and/or animal studies included in this paper. 

\noindent \textbf{Competing Interests} The authors have no competing interests to declare that are relevant to the content of this article. 

\noindent \textbf{Funding} There is no external funding received for this work. 

\noindent \textbf{Availability of data and materials} For data availability, due to the sensitive nature of the research, detailed vulnerability data in browser-based wallets cannot be openly shared. Basic wallet information used in experiments will be available, with detailed vulnerability data provided upon reasonable request for confidentiality and security reasons. 

\end{document}